\documentclass{aastex62}

\graphicspath{{./}{figures/}}

\submitjournal{ApJ}

\shorttitle{He-WD in M3B}
\shortauthors{Cadelano et al.}

\begin{document}

\title{An extremely low-mass He white dwarf orbiting the millisecond pulsar J1342$+$2822B in the globular cluster M3}

\correspondingauthor{Mario Cadelano}
\email{mario.cadelano@unibo.it}

\author[0000-0002-0786-7307]{M. Cadelano}
\affil{Dipartimento di Fisica e Astronomia, Via Gobetti 93/2 I-40129 Bologna, Italy}
\affil{INAF – Osservatorio di Astrofisica e Scienze dello Spazio di Bologna, Via Gobetti 93/3 I-40129 Bologna, Italy}

\author{F. R. Ferraro}
\affiliation{Dipartimento di Fisica e Astronomia, Via Gobetti 93/2 I-40129 Bologna, Italy}
\affil{INAF – Osservatorio di Astrofisica e Scienze dello Spazio di Bologna, Via Gobetti 93/3 I-40129 Bologna, Italy}

\author{A. G. Istrate}
\affiliation{Department of Astrophysics/IMAPP, Radboud University Nijmegen, PO Box 9010, NL-6500 GL Nijmegen, the Netherlands}

\author{C. Pallanca}
\affiliation{Dipartimento di Fisica e Astronomia, Via Gobetti 93/2 I-40129 Bologna, Italy}
\affil{INAF – Osservatorio di Astrofisica e Scienze dello Spazio di Bologna, Via Gobetti 93/3 I-40129 Bologna, Italy}

\author{B. Lanzoni}
\affiliation{Dipartimento di Fisica e Astronomia, Via Gobetti 93/2 I-40129 Bologna, Italy}
\affil{INAF – Osservatorio di Astrofisica e Scienze dello Spazio di Bologna, Via Gobetti 93/3 I-40129 Bologna, Italy}

\author{P. C. C. Freire}
\affiliation{Max-Planck-Institut für Radioastronomie MPIfR, Auf dem Hügel 69, D-53121 Bonn, Germany}

\begin{abstract}

We report on the discovery of the companion star to the millisecond pulsar J1342+2822B in the globular cluster M3. We exploited a combination of near-ultraviolet and optical observations acquired with the {\it Hubble Space Telescope} in order to search for the optical counterparts to the known millisecond pulsars in this cluster. At a position in excellent agreement with that of the radio pulsar J1342$+$2822B (M3B), we have identified a blue and faint object ($m_{F275W}\approx22.45$) that, in the color-magnitude diagram of the cluster, is located in the region of He core white dwarfs. From the comparison of the observed magnitudes with theoretical cooling tracks we have estimated the physical properties of the companion star: it has a mass of only $0.19\pm0.02 \ M_{\odot}$, a surface temperature of $12\pm 1 \cdot 10^3$ K and a cooling age of $1.0^{+0.2}_{-0.3}$~Gyr. Its progenitor was likely a $\sim0.84 \ M_{\odot}$ star and the bulk of the mass-transfer activity occurred during the sub-giant branch phase. The companion mass, combined with the pulsar mass function, implies that this system is observed almost edge-on and that the neutron star has a mass of $1.1\pm0.3 \ M_{\odot}$, in agreement with the typical values measured for recycled neutron stars in these compact binary systems. We have also identified a candidate counterpart to the wide and eccentric binary millisecond pulsar J1342+2822D. It is another white dwarf with a He core and a mass of $0.22\pm0.2 \ M_{\odot}$, implying that the system is observed at a high inclination angle and hosts a typical NS with a mass of $1.3\pm0.3 \ M_{\odot}$. At the moment, the large uncertainty on the radio position of this millisecond pulsar prevents us from robustly concluding that the detected star is its optical counterpart. 

\end{abstract}


\section{Introduction} \label{sec:intro}

Millisecond pulsars (MSPs) are rapidly spinning and highly magnetized neutron stars (NSs), formed through mass and angular momentum accretion from an evolving companion star to a slowly rotating NS \citep{alpar82,stairs04,tauris11}. At the end of the accretion stages, which are commonly observed in low-mass X-ray binaries \citep[e.g.][]{tauris06,papitto13,ferraro15}, the re-accelerated NS is reactivated as a MSP in the radio bands and the companion star is expected to be an exhausted and deeply peeled star, typically a white dwarf (WD) with a He core \citep[e.g.][]{stairs04}, although deviations from this scenario exist \citep[e.g.][]{ferraro01a,ransom05,lynch12,pallanca12,pallanca13b,cadelano15a}.
Globular clusters (GCs) are the ideal factory for the formation of MSPs. In fact, the large stellar densities in the cores of GCs favor dynamical interactions, such as exchange interactions and tidal captures, which can lead to the formation of a large variety of binary systems whose evolution generates stellar exotica like blue stragglers \citep[e.g.][]{ferraro09,ferraro12,ferraro18a}, low-mass X-ray binaries \citep[e.g.][]{pooley03}, cataclysmic variables \citep[e.g.][]{paresce92,ivanova06} and MSPs as well \citep[e.g.][]{ransom05,hessels07,cadelano18}. As a consequence, the number of MSPs per unit mass in the global GC system of the Milky Way turns out to be $\sim 10^3$ times larger than in the Galactic field. 

{To date, 150 MSPs are known in 28 different GCs. Among these clusters, the case of Terzan 5 and 47 Tucanae are particularly worthy of attention: the former is a stellar system known to host $\sim25\%$ of the entire MSP population in GCs \citep{ransom05,cadelano18,andersen18} and proposed to be the remnant of a pristine fragment of the Galactic bulge \citep{ferraro09,lanzoni10,massari14,ferraro16,prager17}; the latter is the second cluster in terms of MSP abundance, hosting 25 systems that allowed studies ranging from binary evolution to the cluster structure and gas content \citep[see e.g.][]{freire01,cadelano15b, riverasandoval15,ridolfi16,freire17,abbate18}}. Therefore the study of MSPs deserves special attention since it allows to study the physical properties of the host stellar system \citep[e.g.][]{freire17,prager17,abbate18} and also to probe binary  and stellar evolution under extreme conditions. This can be done, for example, through the identification of the optical counterparts, whose emission is dominated by the companion stars \citep{ferraro01a,ferraro03a,mucciarelli13,pallanca14,cadelano15b,cadelano17}. In the case of degenerate companions, like WDs or NSs, precise mass measurements of both the components of the binary can be performed directly through timing analysis if relativistic effects are observed. Such values, not only are highly valuable to study in detail the formation and evolution of these systems, but are also of extreme importance to put constraints on the equation of state of ultra-dense matter and to test general relativity \citep[e.g.][]{freire08,demorest10,antoniadis13}. If relativistic effects are not observed, mass measurements can be obtained through the optical identification of the WD companion. In fact, by comparing the observed magnitudes with appropriate WD cooling sequences, the mass of the companion star can be evaluated and such a value, combined with the mass function obtained through radio timing, provides the determination of the NS mass. { Furthermore, if the companion star is bright enough, the radial velocity curve, and thus the mass ratio,  can be determined through spectroscopic observations, together with more precise measurements of the companion mass} \citep[see e.g.][]{ferraro03b,antoniadis12,antoniadis13}. 

M3 (NGC 5272) is a bright and well studied GC located at about 10 kpc from the Sun \citep[see][]{ferraro93,buonanno94,ferraro97a,ferraro97b,ferraro97c,carretta98,rood99,ferraro18b}. Its stellar population is characterized by an intermediate metallicity ($[Fe/H]=-1.5$), a very small reddening $E(B-V)=0.01$ and a large number of variable stars \citep[see][for more information about the cluster properties]{harris10}. \citet{hessels07} reported on the secure identification of 3 MSPs in this cluster plus a fourth candidate. Among these, only two MSPs, namely J1342$+$2822B and J1342$+$2822D (hereafter M3B and M3D) have been clearly detected  in a number of observations large enough to provide the astrometric and orbital properties of the binaries, although the position of M3D is still affected by a large uncertainty in both right ascension and declination. The properties of these two systems, as reported by \citet{hessels07}, are summarized in Table~\ref{tab:MSP}. M3B is a 2.39 ms MSP, located in binary with a $\sim 1.4$ days circular orbit. Its general radio properties suggest that it is a canonical system and therefore it should be in a binary system with a WD companion. On the other hand, M3D has a spin period of 5.45 ms and it is in a very long and slightly eccentric orbit of about 129 days. The vast majority of binary MSPs in GCs have orbital periods shorter than few days, while just a dozen have orbital periods longer than 10 days\footnote{see \url{http://www.naic.edu/~pfreire/GCpsr.html}}. Therefore M3D likely underwent an anomalous evolution; indeed, binaries with orbital periods longer then 10 days are believed to be formed through an exchange interaction between an isolated NS and a primordial cluster binary \citep{hut92,sigurdsson93}. Its eccentricity, which is likely the result of fly-by encounters with other stars \citep{phinney92,phinney93,bagchi09}, is unexpected for systems formed through the canonical low-mass X-ray binary channel and confirms a non-standard evolution for this system.

Here we report on the secure identification of the companion star to M3B and on a candidate companion to M3D. These companions have been discovered through high resolution near-ultraviolet (near-UV) and optical observations. In Section~\ref{sec:data} we present the data-set we used in this work and the data reduction procedures. In Section~\ref{sec:comM3B} we present the identification and characterization of the companion star to M3B, while in Section~\ref{sec:comM3D} we discuss a possible candidate companion to M3D. Finally, in Section~\ref{sec:conc} we draw our conclusions.

\begin{deluxetable*}{lcc}
\tablecolumns{3}
\tablewidth{0pt}
\tablecaption{Main radio timing
parameters for M3B and M3D\tablenotemark{a}, from \citet{hessels07}.\label{tab:MSP}}
\tablehead{\colhead{Parameter} & \colhead{M3B} & \colhead{M3D}}
\startdata
Right ascension, $\alpha$ (J2000)\dotfill  & 13$^{\rm h}\,42^{\rm m}\,11\fs0871(1)$  & 13$^{\rm h}\,42^{\rm m}\,10\fs2(6)$    \\
Declination, $\delta$ (J2000)  \dotfill  & $28^\circ\,22'\,40\farcs141(2)$   & $28^\circ\,22'\,36(14)\arcsec$  \\
Angular offset from cluster center, $\theta_{\perp}$ (\arcsec) \dotfill & 8.4 & 14(7) \\
Spin period, $P$ (ms)  \dotfill  & 2.389420757786(1) &   5.44297516(6) \\
Orbital period, $P_b$ (days) \dotfill & 1.417352298(2) & 128.752(5) \\
Time of ascending node passage, $T_{asc}$ (MJD) \dotfill & 52485.9679712(6) & 52655.38(4) \\
Projected semi-major axis, $x$ (s) \dotfill & 1.875655(2) & 38.524(4)\\
Eccentricity, $e$ \dotfill & ... & 0.0753(5) \\
MSP mass function, $f \ (M_{\odot})$ \dotfill  & 0.003526842(6) & 0.0037031(8) \\
\hline
\enddata
\tablenotetext{a}{Numbers in parentheses are uncertainties in the last digits quoted.}
\end{deluxetable*}

\section{Observations and DAta analysis} \label{sec:data}

This work is based on a archival data-set of deep, high resolution near-UV and optical observations obtained through the UVIS channel of the Wide Field Camera 3 mounted on the Hubble Space Telescope (GO 12605, PI: Piotto). The data-set is composed of 6 images in the F275W {(near-ultraviolet)} filter with exposure time of 415 s, 4 images in the F336W (U) filter with exposure time of 350 s and 4 images in the F438W (B) filter with exposure time of 42 s. All the images {have been obtained on May 15th 2012 during $\sim5$ hours of continual observations and thus they only cover $\sim15\%$ and $\sim0.16\%$ of the orbits of M3B and M3D, respectively}. The images are dithered by few arcseconds each, in order to avoid spurious effects due to bad pixels and to fill the inter-chip gap. The photometric analysis has been performed with {\rm DAOPHOT IV} \citep{stetson87} adopting the so-called ``UV-route'' as described in \citealt{raso17} (see also \citealt{ferraro97b,ferraro01b,ferraro03c,dalessandro18a,dalessandro18b}). As a first step, about 200 stars have been selected in each image in order to model the point spread function, which has been then applied to all the sources detected at more than $5\sigma$ from the background level. Then, we built a master catalog with stars detected in at least half the near-UV (F275W) images. At the corresponding position of these stars, the photometric fit was forced in all the other frames  by using {\rm DAOPHOT/ALLFRAME} \citep{stetson94}. Using such a near-UV master list, the crowding effect due to the presence of giants and turn-off stars is mitigated and several blue stars like blue stragglers and white dwarfs are recovered. Finally, for each star we homogenized the magnitudes estimated in different images, and their weighted mean and standard deviation have been adopted as the star magnitude and its related photometric error. The instrumental magnitudes have been reported to the VEGAMAG system by cross-correlation with the catalog of the ``{\it The Hubble Space Telescope UV Legacy Survey of Galactic globular clusters}'' \citep{piotto15}, obtained from the same data-set used in this work and publicly available. 

The star detector positions have been corrected from geometric distortion following the procedure described by \citet{bellini11}. Then, these positions have been reported to the absolute coordinate system ($\alpha$,$\delta$) using the stars in common with the Gaia DR2 catalog \citep{gaia18}. The coordinate system of this catalog is based on the International Celestial Reference System, which allows an appropriate comparison with the MSP positions derived from timing using solar system ephemerids, being the latter referenced to the same celestial system. The resulting $1\sigma$ astrometric uncertainty is of $\sim 0.1 \arcsec$ in both $\alpha$ and $\delta$.

\section{IDENTIFICATION OF THE COMPANION STAR to M3B} \label{sec:comM3B}

\begin{figure}[t] 
\centering
\includegraphics[scale=0.72]{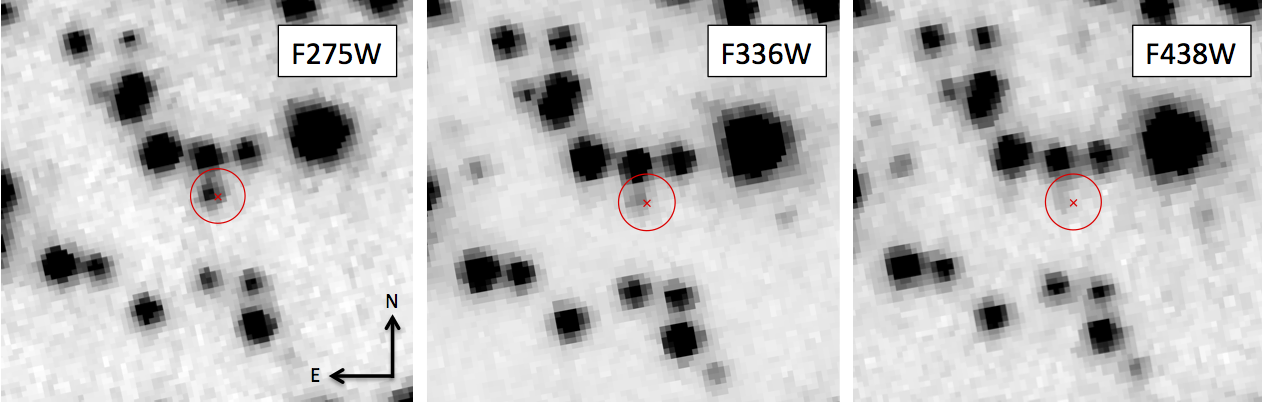}
\caption{$3\arcsec \times 3 \arcsec$ finding chart of the region around the radio position of M3B. The left, middle and right panels are extracted from a F275W, F336W and F438W image, respectively. The red cross is centered on the MSP position and the red circle has a radius $r=0.15\arcsec$. The only star located within this circle is the identified companion to M3B.}
\label{fig:chartB}
\end{figure}

\begin{figure}[h!] 
\centering
\includegraphics[scale=0.49]{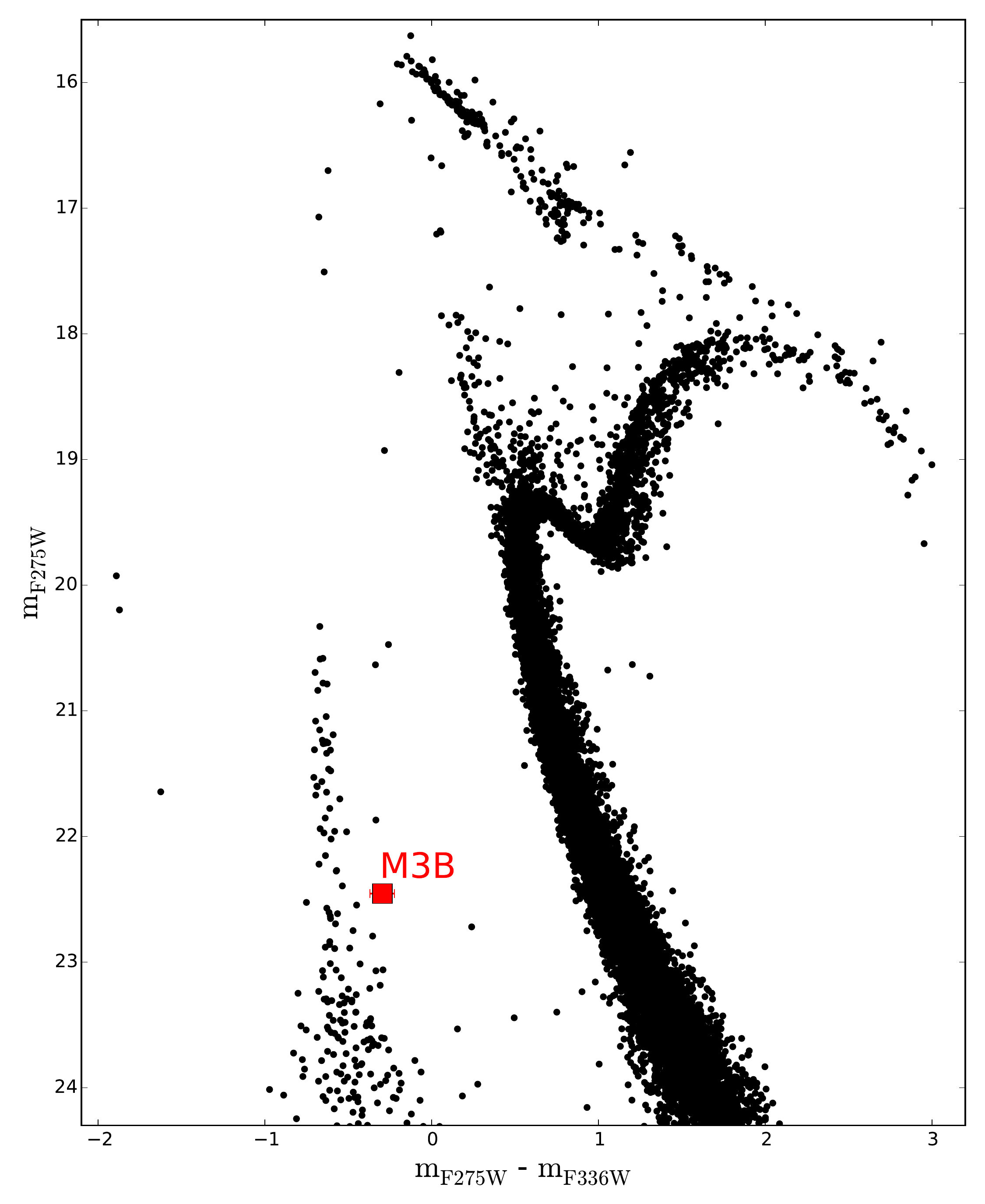}
\caption{($m_{F275W}$, $m_{F275W}-m_{F336W}$) CMD of M3 obtained from the data-set used in this work. The positions of the companion star to M3B is marked with a red square.}
\label{fig:cmd}
\end{figure}

In order to search for the companion star to M3B, we investigated all the stellar sources detected in a $3\arcsec \times 3 \arcsec$ region surrounding the MSP timing position reported in Table~\ref{tab:MSP}. The stellar source closest to the radio position turned out to be a very blue object, located at $\alpha=13^{\rm h}\,42^{\rm m}\,11\fs0881$ and $\delta=28^\circ\,22'\,40\farcs141$, at only $0.01 \arcsec$ from the radio MSP position. The finding chart of this object in the three different filters is reported in Figure~\ref{fig:chartB}. In the color-magnitude diagram (CMD) this star is located along the red side of the WD cooling sequence, in a region where WDs with a He core are expected to be found (see Figure~\ref{fig:cmd}). The combination of the excellent agreement between the radio and the optical position and the peculiar location in the CMD allows us to safely conclude that the identified WD is the companion star to M3B. Its magnitudes in the three filters are the following: $m_{F275W}=22.45\pm0.05$, $m_{F336W}=22.75\pm0.05$ and $m_{F438W}=23.4\pm0.2$. 
Although multiple epoch images are available for each filter, no evidence of photometric variability has been observed for the companion star. While this could be due to the poor and sparse orbital period coverage provided by the available data-set, {we stress that the photometric variability is only rarely observed for degenerate companion stars and is not due to the re-heating of the star by the MSP emitted energy, which is generally negligible, but most likely to pulsations (global stellar oscillations) of the WD itself \citep[e.g.][]{maxted13,kilic15}.}

\begin{figure}[b]
\centering
\includegraphics[scale=0.4]{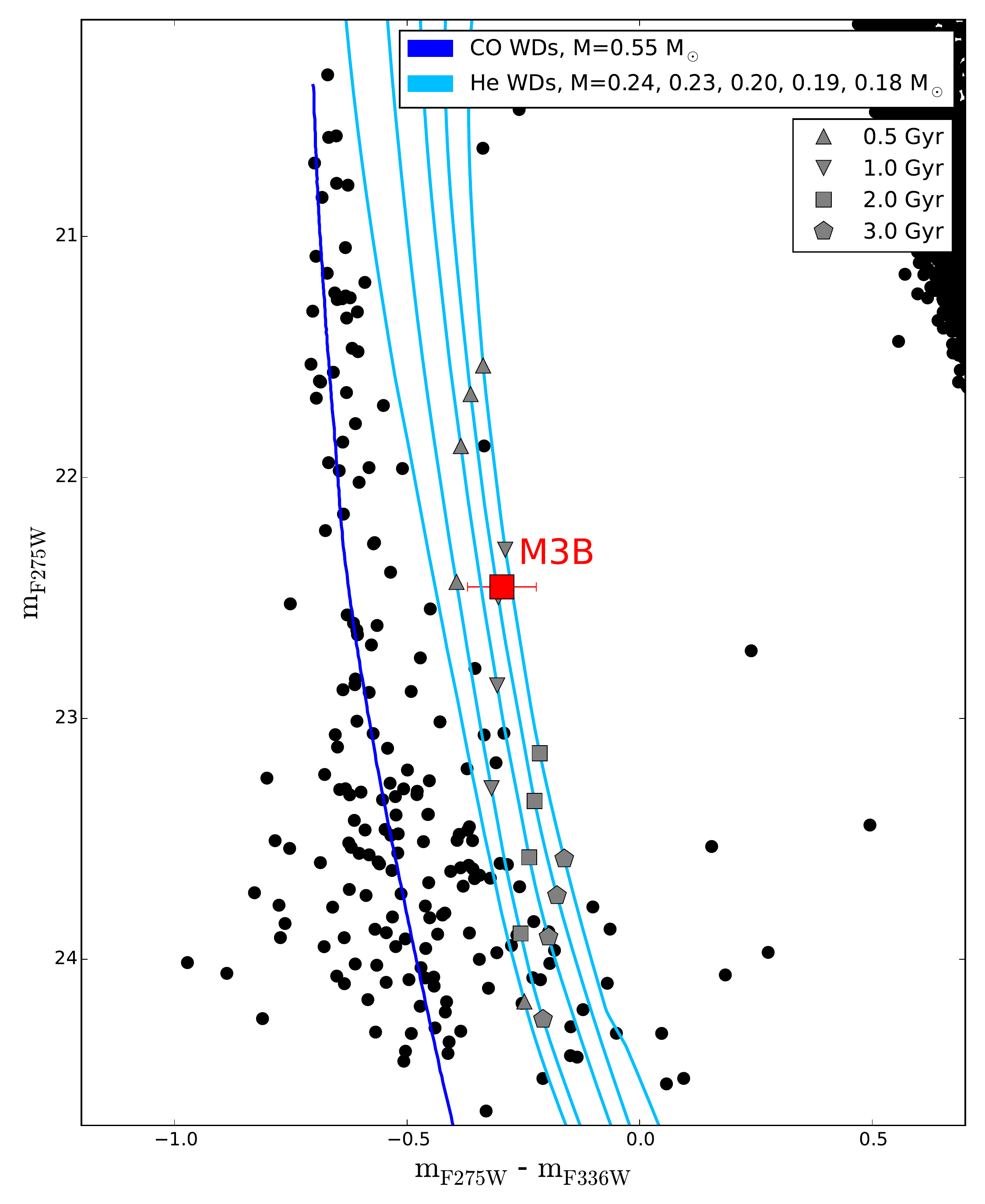}
\caption{Same as in Figure~\ref{fig:cmd} but zoomed into the WD region. As reported in the top legend, the blue curve is a cooling track for $0.55 \, M_{\odot}$ C-O WDs from the {\it BaSTI} database, while the cyan curves are He WD cooling tracks with masses, from left to right, of $0.24 \, M_{\odot}$, $0.23 \, M_{\odot}$, $0.20 \, M_{\odot}$, $0.19 \, M_{\odot}$ and $0.18 \, M_{\odot}$, calculated at the cluster metallicity $(Z\sim0.0005)$ following \citet{istrate14,istrate16}. Points at different cooling ages are highlighted with gray symbols as reported in the secondary legend.}
\label{fig:cmdwd}
\end{figure}

We can now compare the measured magnitudes with those predicted by He WD cooling sequences in order to derive the main physical properties of the companion star. As a first step, we checked the validity of the photometric calibration by comparing the observed sequence of standard WDs with a theoretical cooling model for C-O WDs with masses $M=0.55 \ M_{\odot}$ taken from the {\it BaSTI} database \citep{salaris10}. Absolute magnitudes have been reported to the observed frame by adopting a color-excess $E(B-V)=0.01$, a distance modulus $(m-M)_V=15.07$ \citep{harris10}, and appropriate extinction coefficients for each filter: $A_{F275W}/A_V =1.94$, $A_{F336W}/A_V =1.66$ and $A_{F438W}/A_V =1.33$ \citep{cardelli89,odonnell94}. The corresponding curve is shown in Figure~\ref{fig:cmdwd} and its agreement with the observed WD sequence confirms the accuracy of the adopted photometric calibration and cluster parameters.

In the case of the He WDs, we built a set of theoretical tracks using the new models for extremely low-mass He WD by \citet{istrate14,istrate16}. These models are particularly appropriate for the present case study, since they simulate the entire evolution of systems within the low-mass X-ray binary channel, also including the effects of rotational mixing and element diffusion {and adopting the appropriate cluster metallicity for the secondary stars}. A selection of the obtained cooling tracks is shown in Figure~\ref{fig:cmdwd} (cyan lines).

It is now possible to use the theoretical tracks to constrain the main physical properties of the detected WD such as its mass, surface temperature and cooling age\footnote{The WD cooling age is defined, according to \citet{istrate16}, as the time passed since the proto-WD reached the maximum surface temperature along the evolutionary track.}. To do this, we used an approach similar to that implemented by \citealt{dai17} (see also \citealt{pallanca13a}, \citealt{testa15}). We interpolated all the available models to create a fine grid in the mass range $0.16 \, M_{\odot}$ - $0.35 \, M_{\odot}$, temperature range 9000~K~-~21000~K and cooling age range 0.1~Gyr~-~3~Gyr. Assuming Gaussian photometric errors,  we thus estimated the likelihood of each point of the grid (i.e. of each possible combination of the three parameters) as: 
$$
L = \prod_{j} \frac{exp\left[\frac{-\left(m_j - m_j^{mod}\right)^2}{2\delta_j^2}\right]}{\sqrt{2\pi \delta_j^2}} 
$$
where the index $j$ runs through the three photometric filters F275W, F336W and F438W, $m_j$ and $\delta_j$ are the observed magnitudes and related errors and $m_j^{mod}$ is the magnitude predicted by the models at that point of the grid (see equation 5 in \citealt{dai17}).

\begin{figure}[b] 
\centering
\includegraphics[scale=0.45]{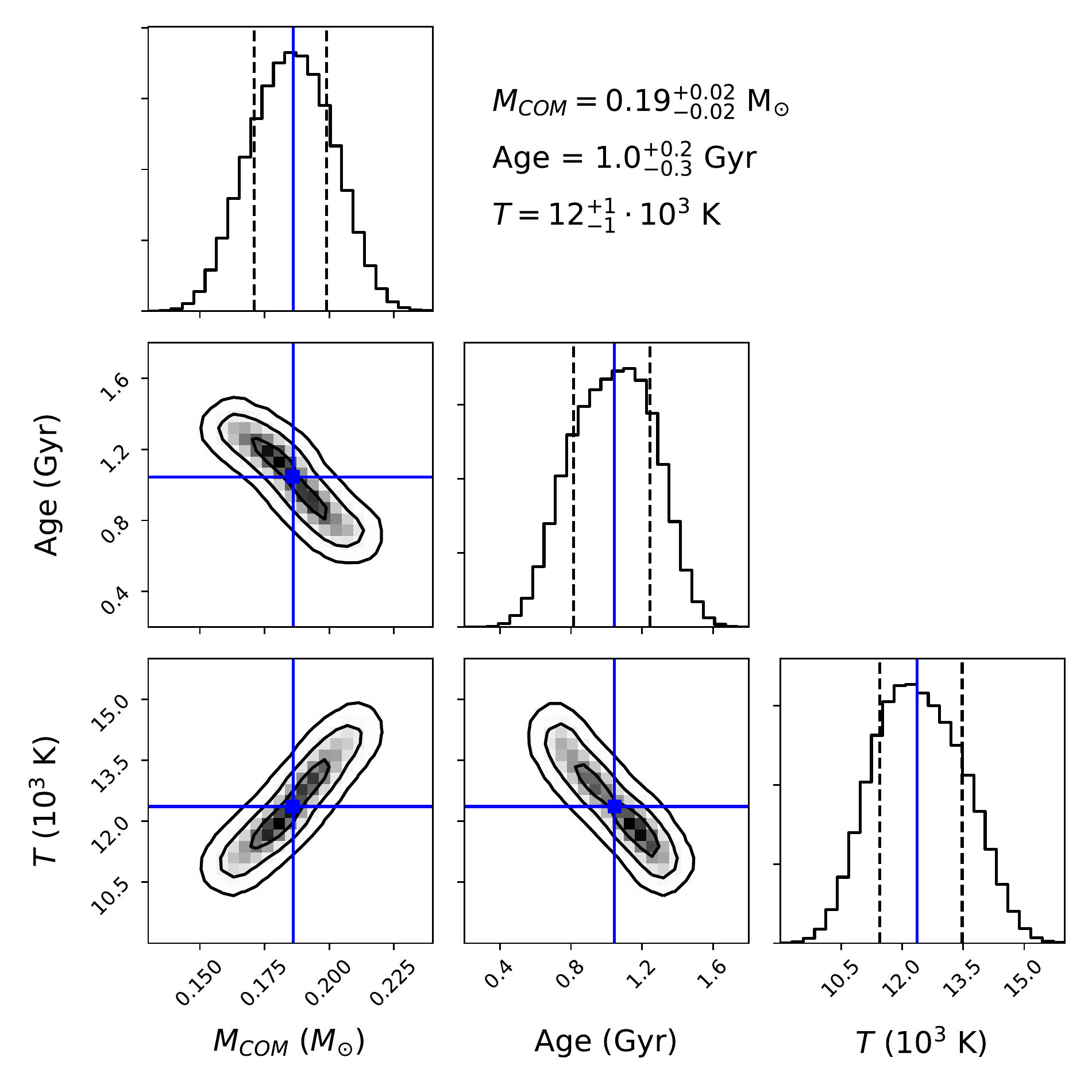}
\caption{Constraints on the mass, cooling age and surface temperature of the companion star to M3B. The 1D histograms show the likelihood marginalized distributions for each of the three parameters and the blue solid and black dashed lines are, respectively, the $50^{th}$, $16^{th}$ and $84^{th}$ percentiles of each distribution, that have been used as estimates of the best-fit value of each parameter and their related uncertainty. The contours in the 2D histograms correspond to $1\sigma$, $2\sigma$ and $3\sigma$ levels and the best values for each parameter are marked with the blue point and lines. The text at the top  reports the derived mass, cooling age and temperature values.}
\label{fig:cornerM3B}
\end{figure}

The marginalized 1D and 2D likelihood distributions are presented in the corner plot\footnote{see \url{https://corner.readthedocs.io/en/latest/} \citep{foreman16}} of Figure~\ref{fig:cornerM3B}. For each parameter, the best-fit value, the lower and the upper uncertainties have been estimated as the $50^{th}$, $16^{th}$ and $84^{th}$ percentile of its likelihood distribution, respectively. We therefore found that the He WD has a mass $M_{COM}~=~0.19~\pm~0.02~ \ M_{\odot}$, a cooling age of $1.0^{+0.2}_{-0.3}$ Gyr and an effective temperature of $T=12\pm 1\cdot10^3$ K. The derived cooling age is consistent with the MSP spin-down age ($>1.1$~Gyr) estimated by \citet{hessels07}.
From the best-fit model we have also inferred that the proto-WD phase\footnote{{ It is the phase following the mass-transfer stage, when the He core contracts at an almost constant luminosity before starting its cooling phase \citep[see][]{istrate14,istrate16}}.} lasted for $1.0^{+0.1}_{-0.5}$ Gyr and that this star is composed of a core with a mass of about $0.18 \ M_{\odot}$ and a thin envelope with mass of about $0.01 \ M_{\odot}$. The total He mass is around $0.187 \ M_{\odot}$, while the H mass is only around $0.003 \ M_{\odot}$. Finally, its central density and  temperature are around $1.3 \cdot 10^{5}$~g/cm$^{-3}$ and $2\cdot 10^{7}$~K, respectively.

\begin{figure}[b!]
\centering
\includegraphics[scale=0.32]{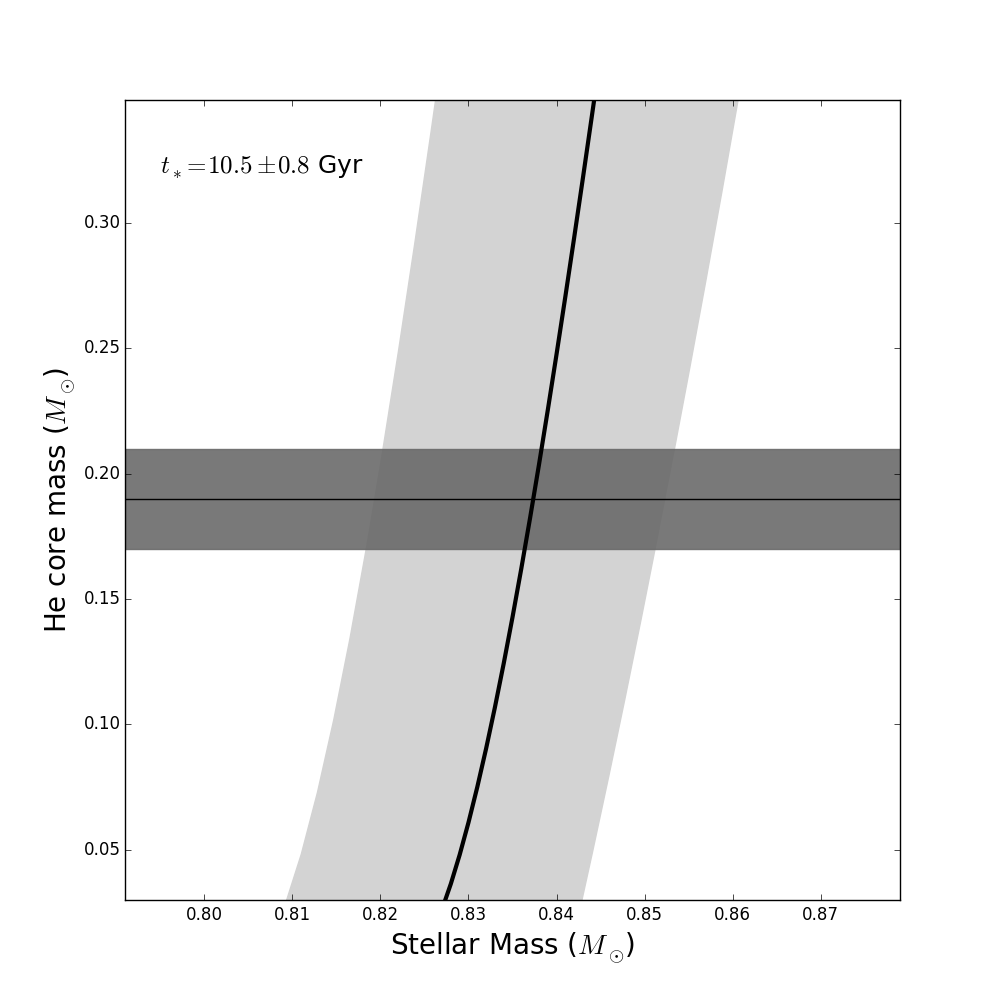}
\includegraphics[scale=0.32]{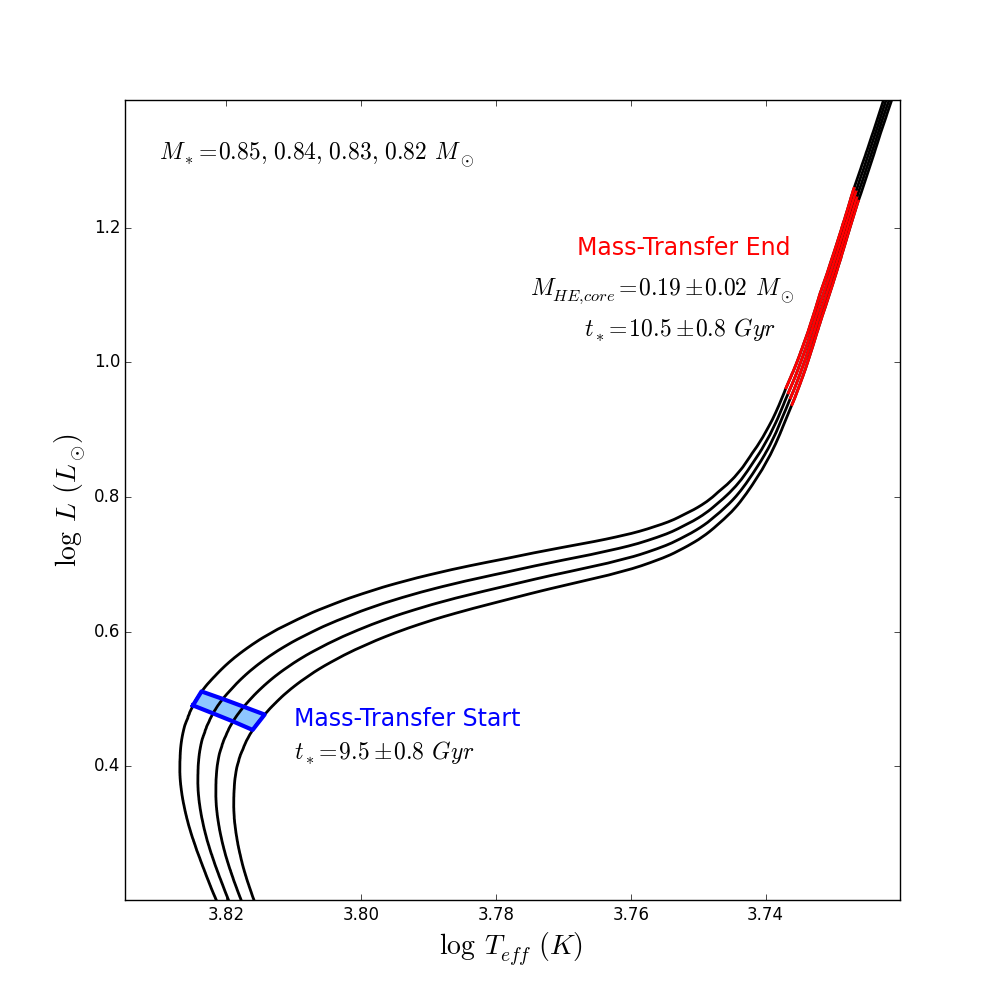}
\caption{{{\it Left Panel:} He core mass as a function of the stellar mass  as predicted by \citet{bressan12,bressan13} evolutionary tracks. The black solid curve and the light gray shaded area represent the values for an age of $10.5\pm0.8$ Gyr. The horizontal line (and dark gray band) marks the mass of the He WD companion to M3B (and its related uncertainty) as derived in Section~\ref{sec:comM3B}. {\it Right Panel:} evolutionary tracks for different stellar masses as reported in the top-left legend. The red region of the tracks highlights the phase where the stars have grown a He core with a mass comparable with the one measured for the companion to M3B at an age of $10.5\pm0.8$ Gyr (corresponding to the mass-transfer end, i.e. the Roche-Lobe detachment phase). The blue shaded area marks the region of the tracks 1 Gyr before the Roche-Lobe detachment, when the mass-transfer probably started.}}
\label{fig:massacore}
\end{figure}

{{The companion star parameters just derived can be used to investigate the physical properties of its progenitor star.} We used the {\rm PARSEC} evolutionary tracks \citep{bressan12,bressan13} to study the evolution of the He core of isolated stars with different masses and ages at the cluster metallicity. Assuming a cluster stellar population age of $12.5\pm0.5$ Gyr \citep{dotter10} and using the derived cooling age and proto-WD phase duration, we can estimate that the mass-transfer stopped (i.e. the Roche-Lobe detachment occurred) when the cluster had an age around $10.5\pm0.8$ Gyr. {Within this age range, only stars in the mass range $0.82 \ M_{\odot} - 0.85 \ M_{\odot}$ have grown a He core with a mass comparable with that derived for the companion to M3B (see left panel of Figure~\ref{fig:massacore}). On the other hand, at an age $t=10.5\pm0.8$ Gyr and at the cluster metallicity, the mass of a star at the main sequence turn-off is $0.81^{+0.01}_{-0.02} \ M_{\odot}$ and stars as massive as $0.82 \ M_{\odot} - 0.85 \ M_{\odot}$ are already evolved toward the red giant branch. Indeed, as shown in the evolutionary tracks plotted in right panel of Figure~\ref{fig:massacore}, stars with mass in the range $0.82 \ M_{\odot} - 0.85 \ M_{\odot}$ have grown a $0.19\pm0.02 \ M_{\odot}$ He core at the base of the red giant branch. By assuming that the mass-transfer phase lasted approximately $\sim1$ Gyr, we can infer that the it started just after the star left the main sequence stage and therefore the bulk of the mass-transfer essentially occurred during the sub-giant branch phase, where the star is expected to expand as a consequence of its canonical evolution and so is able to fill its Roche-Lobe. Therefore the observational properties of the M3B companion measured here appear to be fully consistent with a suitable scenario for the formation of the MSP.}}

\begin{figure}[h]
\centering
\includegraphics[scale=0.6]{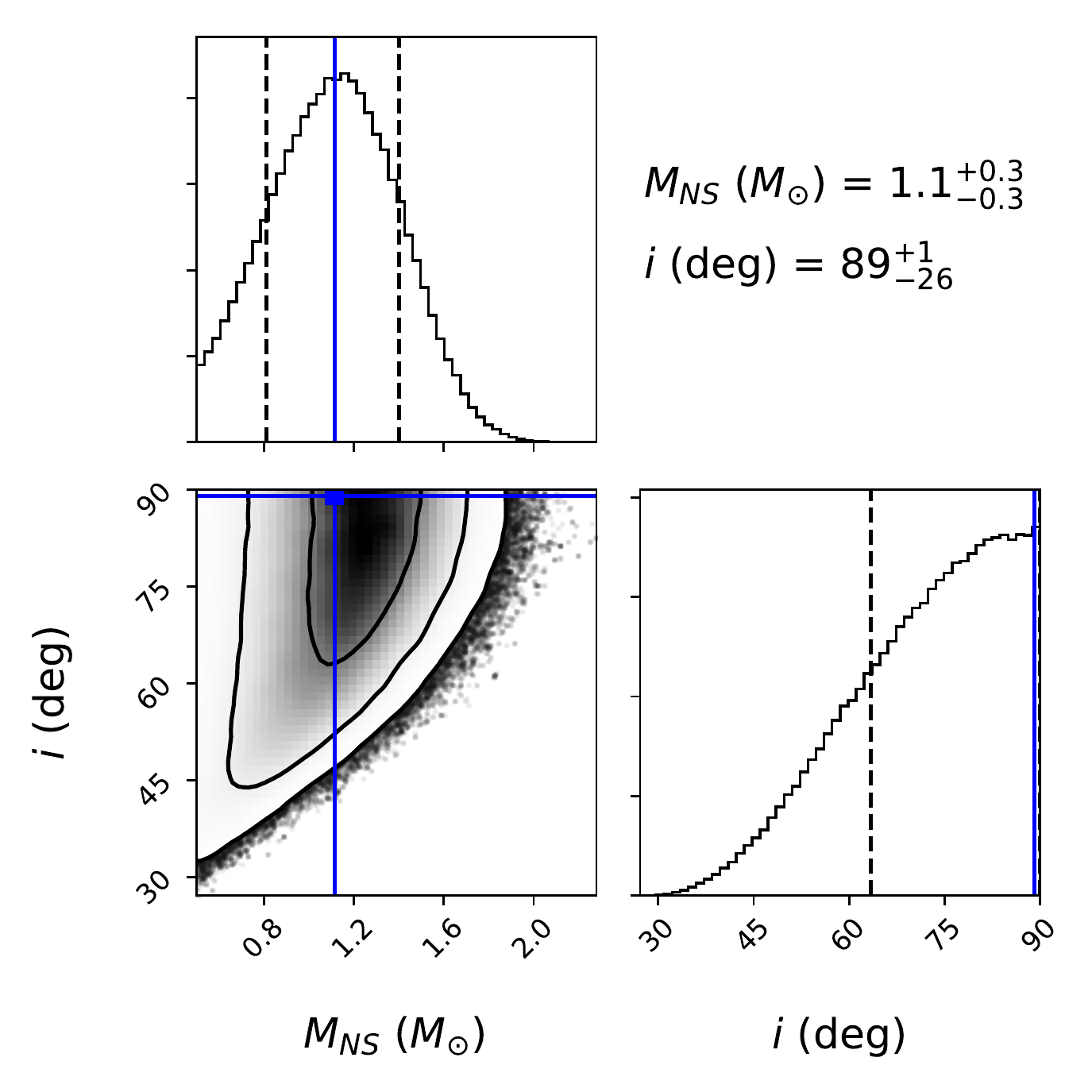}
\caption{Constraints on the mass of the NS and on the orbital inclination angle of M3B. The 1D histograms are the marginalized probability distributions of the two parameters, where the solid blue and black dashed lines are the best values and their related uncertainties (see text). The bottom left panel is the joint 2D posterior probability distribution and the contours corresponds to $1\sigma$, $2\sigma$ and $3\sigma$ confidence levels.}
\label{fig:mpsrB}
\end{figure}

The He WD mass here obtained ($M_{COM}$), combined with the MSP mass function ($f$) derived from radio timing (see Table~\ref{tab:MSP}) can be use to constrain the orbital inclination angle $i$ and, most importantly, the mass of the NS ($M_{NS}$). Indeed these quantities are related by the following equation:
\begin{equation}
f(M_{NS}, M_{COM}, i) = \frac{(M_{COM}\sin{i})^3}{(M_{NS}+M_{COM})^2}
\end{equation}
We used the affine-invariant Markov Chain Monte Carlo (MCMC) ensemble sampler {\rm emcee\footnote{\url{https://emcee.readthedocs.io/en/stable/}}} \citep{foreman13} to constrain $M_{NS}$ and $i$. We set a Gaussian prior on $M_{NS}$, centered at $1.4 \ M_{\odot}$, which is the mass typically measured for NSs in binary MSPs \citep{ozel16}, and with a standard deviation of $0.5 \ M_{\odot}$, {large enough to include the observed NS mass distribution \citep{antoniadis16}}. $M_{NS}$ is sampled in the range $0.5 \ M_{\odot}$ - $2.5 \ M_{\odot}$. On the other hand, we set an uniform prior on the distribution of $\cos{i}$ in the range 0 - 1. The results are presented in Figure~\ref{fig:mpsrB}. The best value for the NS mass is $M_{NS}=1.1\pm 0.3 \ M_{\odot}$, {thus suggesting that this system hosts a low-mass NS, although the value is still compatible with the typical NS masses measured for these kind of systems. Results do not change if wider NS mass distributions are used}. The probability distribution of the inclination angle is shaped as a truncated Gaussian and clearly shows that this system is observed almost edge-on. We therefore assumed that the best value for the inclination angle corresponds to the maximum in the probability distribution and its lower uncertainty to the $16^{th}$ percentile of an identical but symmetric probability distribution. Doing so we have estimated the inclination angle to be $i=89^{+1}_{-26}$ degrees. { All the physical properties of the binary here derived are summarized in Table~\ref{tab:comMSP}.}

\section{A He WD ORBITING M3D TOO?}\label{sec:comM3D}

\begin{figure}[h] 
\centering
\includegraphics[scale=0.16]{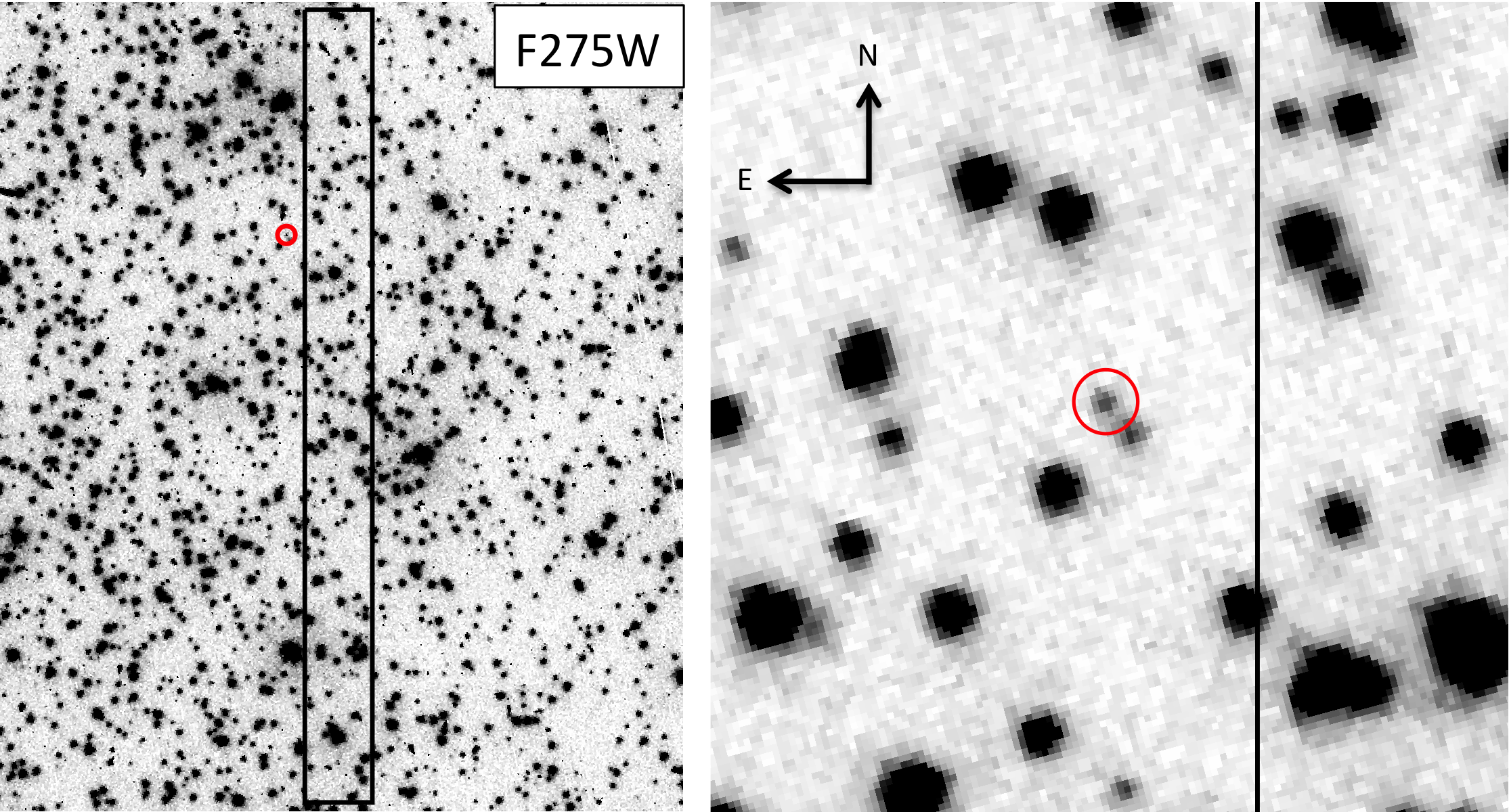}
\caption{Left panel: $15\arcsec \times 18\arcsec$ chart of the region surrounding M3D, obtained from a F275W image. The black box is centered on the MSP position and corresponds to the error box quoted in \citet{hessels07}. The red circle indicates the position of the candidate companion. Right panel: same as in the left panel, but zoomed into the position of the candidate companion.}
\label{fig:chartD}
\end{figure}

\begin{figure}[h!] 
\centering
\includegraphics[scale=0.35]{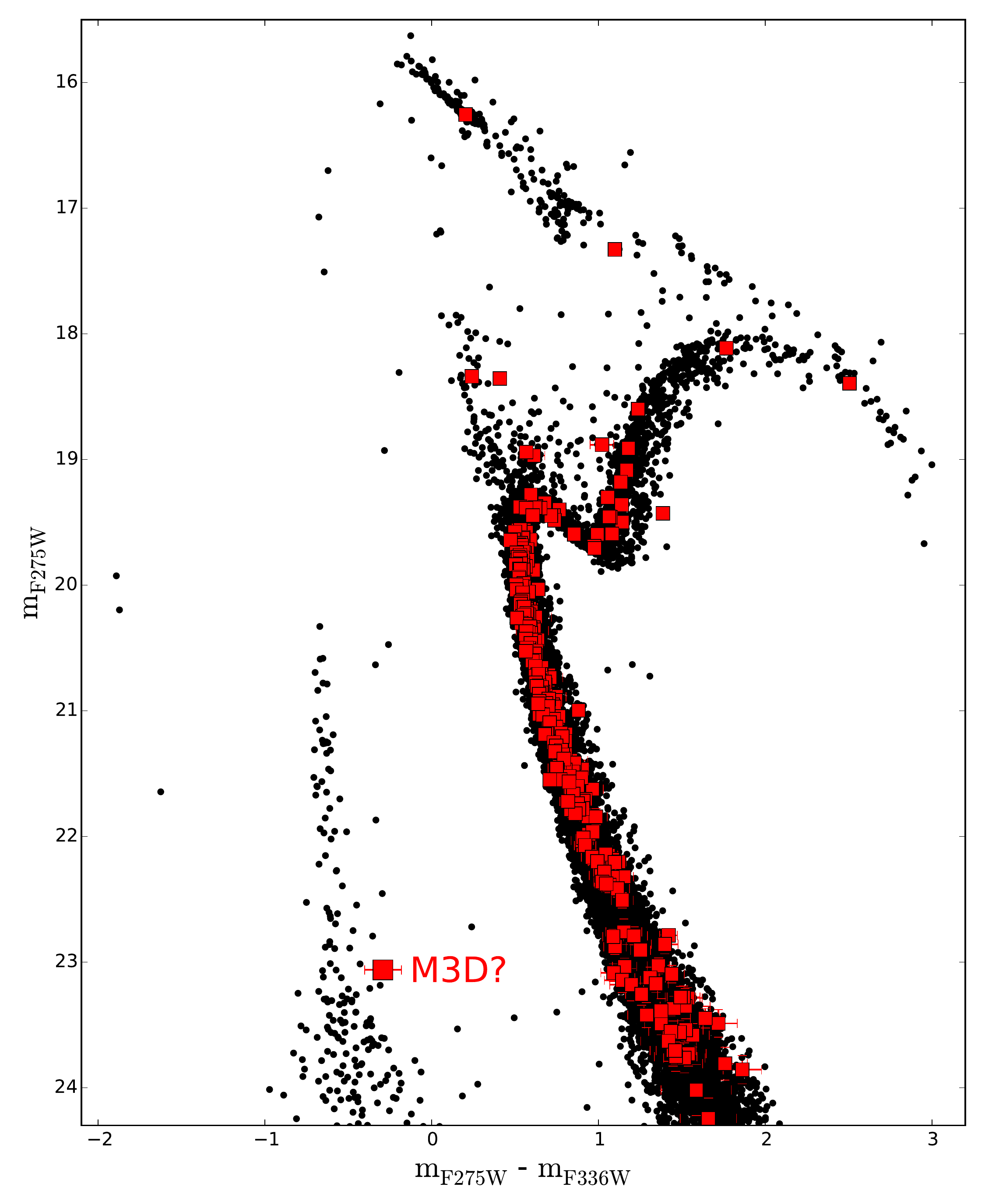}
\includegraphics[scale=0.35]{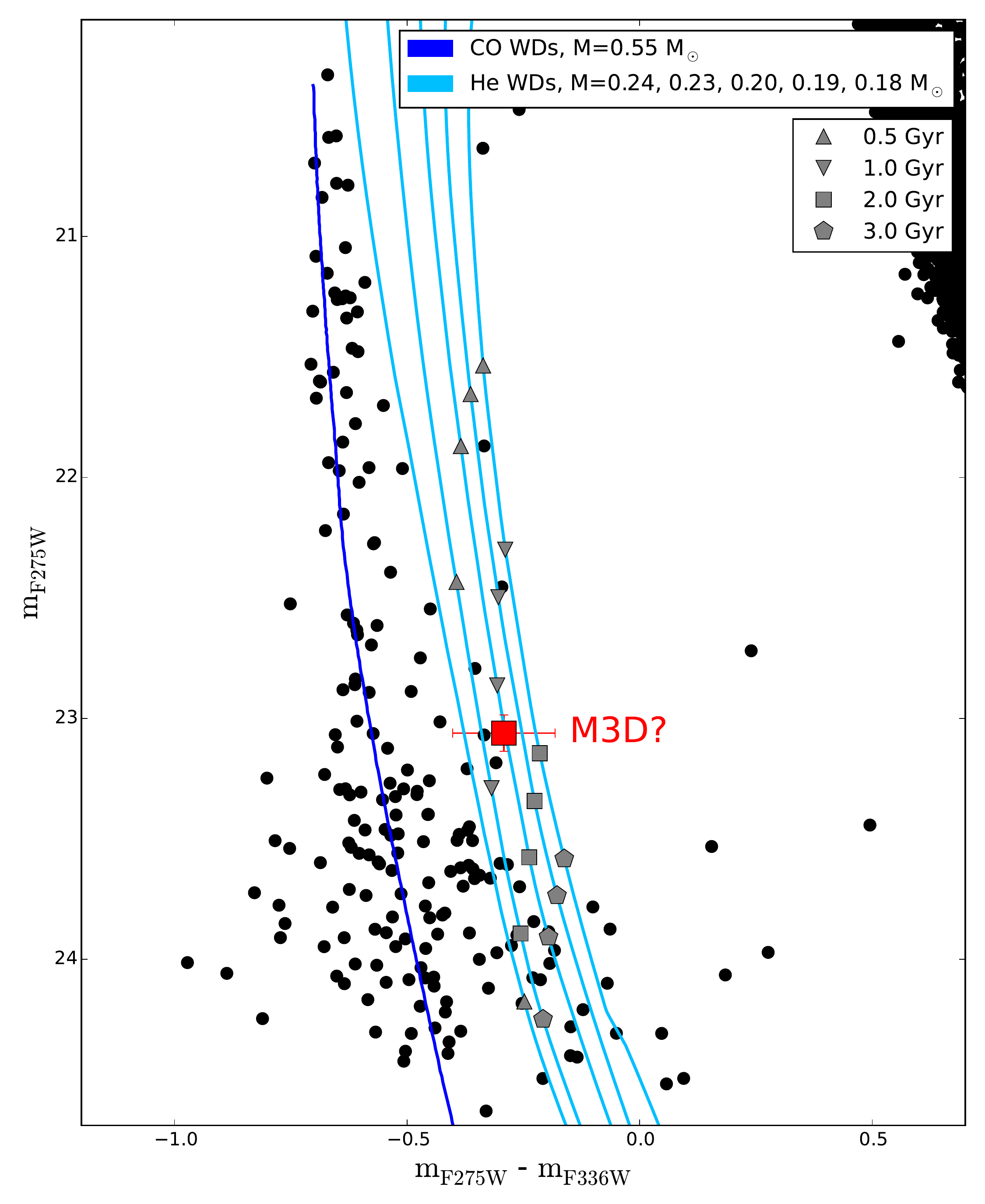}
\caption{{{\it Left panel:} same as in Figure~\ref{fig:cmd}. The position of the candidate companion star to M3D is marked with a large red square, while the positions of all the other stars within the investigated area is marked with smaller red squares. {\it Right panel:} same as in Figure~\ref{fig:cmdwd} but for the candidate companion to M3D.}}
\label{fig:cmdD}
\end{figure}

The timing solution of M3D reported by \citet{hessels07} provides the celestial position of this binary affected by a modest uncertainty ($0.6\arcsec$) in $\alpha$ and a large uncertainty ($14\arcsec$) in $\delta$. Therefore the search for the optical counterpart to this system is challenging. We carefully investigated all the stellar sources within and nearby the position error box: {this region of interest is centered on the best MSP position (see Table~\ref{tab:MSP}) and is as large as the $3\sigma$ uncertainty in $\alpha$ and $2\sigma$ uncertainty in $\delta$. We only found a possible candidate in a position marginally compatible with the radio one (see Figure~\ref{fig:chartD} and Figure~\ref{fig:cmdD}). In fact, all the stars detected within the investigated area likely belong to the canonical evolutionary sequences but one source located again on the red side of the WD cooling sequence, in a position compatible with that of the He WDs}. This candidate counterpart is located at $\alpha=13^{\rm h}\,42^{\rm m}\,10\fs3041$ and $\delta=28^\circ\,22'\,44\farcs786$. The right ascension is shifted by $1.5\arcsec$ from the radio value, which is larger than the combined radio uncertainty and optical astrometric precision.  Its observed magnitudes are: $m_{F275W}=23.06\pm0.08$, $m_{F336W}=23.35\pm0.08$ and $m_{F438W}=24.1\pm0.3$.
Following the same method used for M3B and presented in Section~\ref{sec:comM3B}, we found that this star has a mass $M_{COM}=0.22 \pm 0.02 \ M_{\odot}$, a cooling age of $1.1^{+0.7}_{-0.6}$ Gyr and an effective temperature of $13\pm 2 \cdot 10^{3}$ K. Using the MSP mass function (Table~\ref{tab:MSP}) we estimated the NS mass to be $M_{NS}=1.3 \pm 0.3 \ M_{\odot}$ and the orbital inclination angle to be $i=72\pm16$ degrees (see Table~\ref{tab:comMSP}). Again, the $M_{NS}$ value would be in agreement with the typical values observed for binary MSPs.
We tried to establish phase-connection of the MSP timing solution using the available radio data and fixing the MSP position to that of the candidate counterpart. Unfortunately, this did not lead to an improvement of the timing solution obtained by \citet{hessels07}. 
Given all this, we cannot solidly confirm that the detected He WD is the optical counterpart to M3D, {which could be still under the detection threshold. Furthermore, the peculiar evolution that this system likely underwent implies that the companion star might not be a classical He WD. However it is worth noting that a He WD has been identified orbiting B1620$-$26 in M4, the only similar MSP with an optical counterpart to date \citep{sigurdsson03}.} Future radio observations and timing analysis are needed to improve the position measurement of this object.

\begin{deluxetable*}{lcc}
\tablecolumns{3}
\tablewidth{0pt}
\tablecaption{Physical properties of the binaries M3B and M3D as derived in Section~\ref{sec:comM3B} and ~\ref{sec:comM3D}. \label{tab:comMSP}}
\tablehead{\colhead{Parameter} & \colhead{M3B} & \colhead{M3D?}}
\startdata
Companion mass, $M_{COM} \ (M_{\odot})$\dotfill  & $0.19\pm0.02$ & $0.22\pm0.02$    \\
Effective temperature, $T$ ($10^3$ K) \dotfill & $12\pm1$ & $13\pm2$ \\
Cooling age, Age (Gyr) \dotfill & $1.0^{+0.2}_{-0.3}$ & $1.1^{+0.7}_{-0.6}$ \\
Neutron star mass, $M_{NS} \ (M_{\odot})$\dotfill  & $1.1\pm0.3$ & $1.3\pm0.3$    \\
Inclination angle, $i$ (deg) \dotfill & $89^{+1}_{-26}$ & $72\pm16$ \\
\hline
\enddata
\end{deluxetable*} 

\section{CONCLUSIONS}\label{sec:conc}

We used deep and high resolution images obtained at near-UV and optical wavelengths to search for the companion stars to the binary MSPs in the GC M3. By exploiting the ``UV route'' to dilute the crowding issues in the cluster center and increase the sensitivity to blue/hot stars, we have been able to firmly identify the companion star to the canonical MSP M3B and find a candidate counterpart to the anomalous (long period and mild eccentricity) system M3D.
The companion star to M3B turned out to be a WD with a He core, as expected from the canonical formation scenario. Interestingly, despite the fact that this cluster hosts a stellar population with an intermediate metallicity, the companion is an extremely low-mass object, with a mass of about $0.19 \ M_{\odot}$. Indeed, the lower is the metal content of the secondary stars, the larger is expected to be the minimum mass of the WD remnants. This is due to the fact that low metallicity stars have shorter evolutionary timescales and smaller radii and therefore their Roche Lobe is filled (i.e. the mass-transfer starts) in a more advanced stage of their evolution, when the He core is grown more massive than it would have for a more metal-rich system \citep{istrate16}. Our mass measurement is indeed close to the minimum possible mass produced by the adopted models at the cluster metallicity. This stresses the importance of the study of extremely low-mass WD systems, with special care to the physical processes occurring during the evolution. In fact, the models we used take into account effects like rotational mixing and element diffusion. Models not including these effects are characterized by bluer cooling tracks that would have not been able to reproduce the observed magnitudes of the companion or, in the case they do, the resulting companion mass would have been so small that the corresponding NS mass becomes unreasonable (i.e. $<1 \ M_{\odot}$). {We have also shown that the progenitor of this WD was likely a $\sim0.83 \ M_{\odot}$ star which filled its Roche-Lobe after leaving the main sequence, thus implying that the bulk of the mass-transfer activity occurred during the sub-giant branch phase. All the derived physical properties of the companion star, combined with the information obtained through radio timing, allowed to infer that this binary is observed almost edge-on and probably hosts a low-mass NS with a mass around $\sim 1.1 \ M_{\odot}$.}

In the case of the candidate companion star to M3D, we have identified a He WD at a position marginally compatible with the highly uncertain radio one. Therefore the association between this degenerate object and the MSP cannot be confirmed yet. This WD is again a low-mass object with a mass around $0.22 \ M_{\odot}$ and a cooling age of $\sim1$~Gyr.

{\citet{tauris99} discussed the correlation between the masses of the WD companion stars and the orbital periods of the binary MSPs (the so-called ``TS99'' relation; see also \citealt{istrate14} for a more recent investigation on this). According to this relation, the $\sim1.4$~days orbit of M3B implies a companion star with a mass of $\sim0.21 \ M_{\odot}$, fully compatible, within the uncertainties, with the mass value derived in this work. On the other hand, the long $\sim 129$ days orbit of M3D implies a companion star with a mass around $\sim 0.35 \ M_{\odot}$, a value significantly larger than the one here derived for its candidate companion star. This could suggests that the proposed optical counterpart is not the real companion star, even though its mass is compatible with the measured MSP mass function if a typical NS mass is assumed. If this WD is truly the companion star to M3D, the discrepancy between the measured mass and the one predicted by the TS99 relation could be explained by the formation of this system through an exchange interaction occurred after the NS recycling phase. Such an exchange, however, is expected to produce highly eccentric binaries \citep[see][]{prince91,freire04,lynch12}, while only a mild eccentricity is measured for this system.}

\acknowledgments
{We thank the anonymous referee for the careful reading of the manuscript}. This research is part of the scientific project Cosmic-Lab at the Department of Physics and Astronomy at Bologna University. 

\vspace{5mm}
\facilities{HST(WFC3)}
\software{DAOPHOT \citep{stetson87}, {\rm emcee} \citep{foreman13}, {\rm corner.py} \citep{foreman16}}

\end{document}